# An Analysis of the Cloud Computing Security Problem

Mohamed Al Morsy, John Grundy and Ingo Müller
*Computer Science & Software Engineering, Faculty of Information & Communication Technologies*
*Swinburne University of Technology, Hawthorn, Victoria, Australia*
{malmorsy, jgrundy, imueller}@ swin.edu.au

*Abstract* — Cloud computing is a new computational paradigm that offers an innovative business model for organizations to adopt IT without upfront investment. Despite the potential gains achieved from the cloud computing, the model security is still questionable which impacts the cloud model adoption. The security problem becomes more complicated under the cloud model as new dimensions have entered into the problem scope related to the model architecture, multi-tenancy, elasticity, and layers dependency stack. In this paper we introduce a detailed analysis of the cloud security problem. We investigated the problem from the cloud architecture perspective, the cloud offered characteristics perspective, the cloud stakeholders' perspective, and the cloud service delivery models perspective. Based on this analysis we derive a detailed specification of the cloud security problem and key features that should be covered by any proposed security solution.

## I. INTRODUCTION

Cloud computing provides the next generation of internet-based, highly scalable distributed computing systems in which computational resources are offered 'as a service'. The most widely used definition of the cloud computing model is introduced by NIST [1] as "*a model for enabling convenient, on-demand network access to a shared pool of configurable computing resources (e.g., networks, servers, storage, applications, and services) that can be rapidly provisioned and released with minimal management effort or service provider interaction.*". Multi-tenancy and elasticity are two key characteristics of the cloud model. Multi-Tenancy enables sharing the same service instance among different tenants. Elasticity enables scaling up and down resources allocated to a service based on the current service demands. Both characteristics focus on improving resource utilization, cost and service availability.

The cloud model has motivated industry and academia to adopt cloud computing to host a wide spectrum of applications ranging from high computationally intensive applications down to light weight services. The model is also well-suited for small and medium businesses because it helps  adopting IT without upfront investments in infrastructure, software licenses and other relevant requirements. Moreover, Governments become more interested in the possibilities of using cloud computing to reduce IT costs and increase capabilities and reachability of their delivered services.

According to a Gartner survey [2] on cloud computing revenues, the cloud market  was worth USD 58.6B in 2009 , is expected to be USD 68B in 2010 and will reach USD 148B by 2014. These revenues imply that cloud computing is a promising platform. On the other hand, it increases the attackers' interest in finding existing vulnerabilities in the model.

Despite the potential benefits and revenues that could be gained from the cloud computing model, the model still has a lot of open issues that impact the model creditability and pervasiveness. Vendor lock-in, multi-tenancy and isolation, data management, service portability, elasticity engines, SLA management, and cloud security are well known open research problems in the cloud computing model.

From the cloud consumers' perspective, security is the major concern that hampers the adoption of the cloud computing model [3] because:

- Enterprises outsource security management to a third party that hosts their IT assets (loss of control).
- Co-existence of assets of different tenants in the same location and using the same instance of the service while being unaware of the strength of security controls used.
- The lack of security guarantees in the SLAs between the cloud consumers and the cloud providers.
- Hosting this set of valuable assets on publicly available infrastructure increases the probability of attacks.

From the cloud providers' perspective, security requires a lot of expenditures (security solutions' licenses), resources (security is a resource consuming task), and is a difficult problem to master (as we discuss later). But skipping security from the cloud computing model roadmap will violate the expected revenues as explained above. So cloud providers have to understand consumers' concerns and seek out new security solutions that resolve such concerns.

In this paper we analyze existing challenges and issues involved in the cloud computing security problem. We group these issues into architecture-related issues, service delivery model-related issues, cloud characteristic-related issues, and cloud stakeholder-related issues. Our objective is to identify the weak points in the cloud model. We present a detailed analysis for each weakness to  highlight their root causes. This will help cloud providers and security vendors to have a better understanding of the problem. It also helps researchers being aware of the existing problem dimensions and gaps.

Our paper is organized as follows. In section II, we explore previous efforts in defining cloud security problems and challenges. Sections III to VII explore the cloud computing security problem from different perspectives. Section VIII discusses the key security enablers in the cloud model. Section IX summarizes our conclusions and what we believe are the key dimensions that should be covered by any cloud security solution. Finally, in section X we discuss the future work focusing on one of the discussed security enablers (cloud security management).

## II. LITERATURE REVIEW

Cloud computing security challenges and issues discussed by various researchers. The Cloud Computing Use Cases group [4] discusses the different use case scenarios and related requirements that may exist in the cloud computing model. They consider use cases from different perspectives including customers, developers and security engineers. ENISA [5] investigated the different security risks related to adopting cloud computing along with the affected assets, the risks likelihood, impacts, and vulnerabilities in cloud computing that may lead to such risks. Similar efforts discussed in "Top Threats to Cloud Computing" by CSA [6]. Balachandra et al [7] discuss the security SLA's specifications and objectives related to data locations, segregation and data recovery. Kresimir et al [8] discuss high level security concerns in the cloud computing model such as data integrity, payment, and privacy of sensitive information. Kresimir discussed different security management standards such as ITIL, ISO/IEC 27001 and Open Virtualization Format (OVF). Meiko et al [9] discuss the technical security issues arising from adopting the cloud computing model such as XML-attacks, Browsers' related attacks, and flooding attacks. Bernd et al [10] discuss the security vulnerabilities existing in the cloud platform. The authors grouped the possible vulnerabilities into technology-related, cloud characteristics -related, security controls- related. Subashini et al [11] discuss the security challenges of the cloud service delivery model, focusing on the SaaS model. CSA [6] discusses critical areas of cloud computing. They deliver a set of best practices for the cloud provider, consumers and security vendors to follow in each domain. CSA published a set of detailed reports discussing for some of these domains.

In our research we did a deep investigation in the cloud model to identify the root causes and key participating dimensions in such security issues/problems discussed by the previous work. This will help better to understand the problem and deliver solutions.

## III. THE CLOUD COMPUTING ARCHITECTURE AND SECURITY IMPLICATIONS

The Cloud Computing model has three service delivery models and main three deployment models [1]. The deployment models are: (1) **Private cloud**: a cloud platform is dedicated for specific organization, (2) **Public cloud**: a cloud platform available to public users to register and use the available infrastructure, and (3) **Hybrid cloud**: a private cloud that can extend to use resources in public clouds. Public clouds are the most vulnerable deployment model because they are available for public users to host their services who may be malicious users.

The cloud service delivery models, as in figure1, include:
- **Infrastructure-as-a-service (IaaS)**: where cloud providers deliver computation resources, storage and network as an internet-based services. This service model is based on the virtualization technology. Amazon EC2 is the most familiar IaaS provider.
- **Platform-as-a-service (PaaS):** where cloud providers deliver platforms, tools and other business services that enable customers to develop, deploy, and manage their own applications, *without installing any of these platforms or support tools on their local machines*. The PaaS model may be hosted on top of IaaS model or on top of the cloud infrastructures directly. Google Apps and Microsoft Windows Azure are the most known PaaS.
- **Software-as-a-service (SaaS)**: where cloud providers deliver applications hosted on the cloud infrastructure as internet-based service for end users, *without requiring installing the applications on the customers' computers*. This model may be hosted on top of PaaS, IaaS or directly hosted on cloud infrastructure. SalesForce CRM is an example of the SaaS provider.

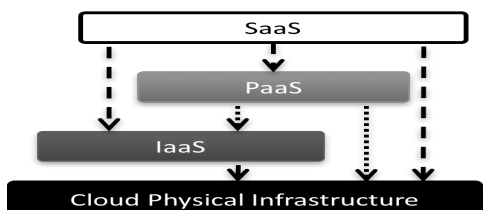

**Figure 1: cloud service delivery models**

Each service delivery model has different possible implementations, as in figure 1, which complicates the development of standard security model for each service delivery model. Moreover, these service delivery models may coexist in one cloud platform leading to further complication of the security management process.

## IV. CLOUD COMPUTING CHARACTERSTICS AND SECURITY IMPLICATIONS

To achieve efficient utilization of resources, cloud providers need to increase their resource utilization while decreasing cost. At the same time consumers need to use resources as far as needed while being able to increase or decrease resources consumption based on actual demands. The cloud computing model meets such needs via a win-win solution by delivering two key characteristics: ***multi-tenancy*** and ***elasticity***. Both characteristics turn out to have serious implications on the cloud model security.

Multi-tenancy implies sharing of computational resources, storage, services, and applications with other tenants. Multi-tenancy has different realization approaches as shown in figure 2. In approach 1, each tenant has their own dedicated instance with their own ***customizations*** *(customization may include special development to meet customer needs)*. In approach 2, each tenant uses a dedicated instance, *like approach 1,* while all instances are the same but with different *configurations (adjustment of application parameters or interfaces)*. In approach 3, all tenants share the same instance with runtime configuration (the application is divided into core application component and extra components that are loaded based on the current tenant requests – similar to SalesForce.com). In approach 4 tenants are directed to a load balancer that redirects tenants requests to a suitable instance based on current instances load. Approaches 3 and 4 are the most risky as tenants are coexisting on the same process in memory and hardware. This sharing of resources violates the confidentiality of tenants' IT assets which leads to the need for secure multi-tenancy. To deliver *secure multi-tenancy* there should be *isolation* among tenants' data (at rest, processing and transition) and *location transparency* where tenants have no knowledge or control over the specific location of their resources (may have high level control on data location such as country or region level), to avoid planned attacks that attempt to co-locate with the victim assets [12]. In *IaaS*, isolation should consider VMs' storage, processing, memory, cache memories, and networks. In *PaaS,* isolation should cover isolatation among running services and APIs' calls. In *SaaS,* isolation should isolate among transactions carried out on the same instance by different tenants and tenants' data.

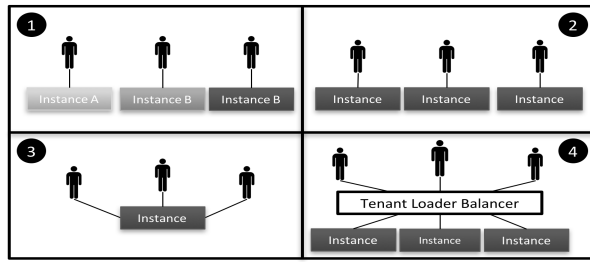

Figure 2: Multi-tenancy approaches [13]

Elasticity implies being able to scale up or down resources assigned to services based on the current demand. Scaling up and down of tenant's resources gives the opportunity to other tenants to use the tenant previously assigned resources. This may lead to confidentiality issues. For example, tenant A scaled down so it releases resources, these resources are now assigned to tenant B who in turn use it to deduce the previous contents of tenant A (similar to lag problem between DNS and DNS cache). Moreover, Elasticity includes a service placement engine that maintains a list of the available resources from the provider's offered resources pool. This list is used to allocate resources to services. Such placement engines should incorporate cloud consumers' security and legal requirements such as avoid placing competitors services on the same server, data location should be within the tenants' country boundaries. Placement engines may include a migration strategy where services are migrated from physical host to another or from cloud to another in order to meet demands and efficient utilization of the resources. This migration strategy should take into account the same security constraints. Furthermore, security requirements defined by service consumers should be migrated with the service and initiates a process to enforce security requirements on the new environment, as defined by cloud consumers, and updates the current cloud security model.

V. CLOUD COMPUTING'S DEEP DEPENDENNCIES STACK

The cloud computing model depends on a deep stack of dependent layers of objects (VMs, APIs, Services and Applications) where the functionality and security of a higher layer depends on the lower ones. The **IaaS** model covers cloud physical infrastructure layer (storage, networks and servers), virtualization layer (hypervisors), and virtualized resources layer (VMs, virtual storage, virtual networks). The **PaaS** model covers the platform layers (such as application servers, web servers, IDEs, and other tools), and APIs and Services layers. The **PaaS** layer depends on the virtualization of resources as delivered by IaaS. The **SaaS** model covers applications and services offered as a service for end users, as shown in figure 3. The **SaaS** layer depends on a layer of platforms to host the services and a layer of virtualization to optimize resources utilization when delivering services to multi-tenant.

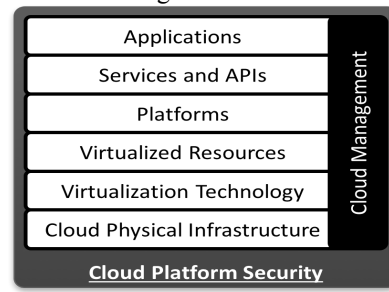

Figure 3: Cloud computing model layers

This deep dependency stack of cloud objects complicates the cloud security problem as the security of each object/layer depends on the security of the lower objects/layers. Furthermore, any breach to any cloud objects will impact the security of the whole cloud platform. Each cloud layer/object has a set of security requirements and vulnerabilities so it requires a set of security controls to deliver secured service. This results in a huge number of security controls that needs to be managed. Moreover, managing such heterogeneous security controls to meet security needs is a complex task, taking into account conflicts among the security requirements and among security controls at each layer. This may result in an inconsistent security model. Hence, a unified security control management module is required. This module should coordinate and integrate among the various layers' security controls based on security needs.

VI. CLOUD COMPUTING STAKEHOLDERS AND SECURITY IMPLICATIONS

The cloud computing model has different involved stakeholders: **cloud provider** (an entity that delivers infrastructures to the cloud consumers), **service provider** (an entity that uses the cloud infrastructure to deliver applications/services to end users), and **service consumer** (an entity that uses services hosted on the cloud infrastructure). Each stakeholder has their own security management systems/processes and each one has their own expectations (requirements) and capabilities (delivered) from/to other stakeholders. This leads to: (1) A set of security requirements defined on a service by different tenants that may conflict with each other. So security configurations of each service should be maintained and enforced on the service instances level and at runtime taking into account the possibility of changing requirements based on current consumers' needs to mitigate new risks; (2) Providers and consumers need to negotiate and agree on the applied security properties. However, no standard security specification notations are available that can be used by the cloud stakeholders to represent and reason about their offered/required security properties; and (3) Each stakeholder has their own security management processes used to define their assets, expected risks and their impacts, and how to mitigate such risks. Adopting cloud model results in losing control from both involved parties, including cloud providers (who are not aware of the contents and security requirements of services hosted on their infrastructures) and cloud consumers (who are not able to control neither on their assets security nor on other services sharing the same resources). Security SLA management frameworks represent part of the solution related to security properties specification, enforcement and monitoring. However, SLAs still don't cover security attributes in their specifications [14]. Moreover, SLAs are high level contracts where the details of the security policies and security control and how to change at runtime are not included.

On the other side, cloud providers are not able to deliver efficient and effective security controls because they are not aware of the hosted services' architectures. Furthermore, cloud providers are faced with a lot of changes to security requirements while having a variety of security controls deployed that need to be updated. This further complicates the cloud providers' security administrators' tasks. Transparency of what security is enforced, what risks exist, and what breaches occur on the cloud platform and the hosted services must exist among cloud providers and consumers. This is what is called "trust but verify" [15], where cloud consumers should trust in their providers meanwhile cloud providers should deliver tools to help consumers to verify and monitor security enforcements.

## VII. CLOUD COMPUTING SERVICE DELIVERY MODELS AND SECURITY IMPLICATIONS

We summarize the key security issues/vulnerabilities in each service delivery model. Some of these issues are the responsibility of cloud providers while others are the responsibility of cloud consumers.

### A. IaaS Issues

*VM security* – securing the VM operating systems and workloads from common security threats that affect traditional physical servers, such as malware and viruses, using traditional or cloud-oriented security solutions. The VM's security is the responsibility of cloud consumers. Each cloud consumer can use their own security controls based on their needs, expected risk level, and their own security management process.

*Securing VM images repository* - unlike physical servers VMs are still under risk even when they are offline. VM images can be compromised by injecting malicious codes in the VM file or even stole the VM file itself. Secured VM images repository is the responsibilities of the cloud providers. Another issue related to VM templates is that such templates may retain the original owner information which may be used by a new consumer.

*Virtual network security* - sharing of network infrastructure among different tenants within the same server (using vSwitch) or in the physical networks will increase the possibility to exploit vulnerabilities in DNS servers, DHCP, IP protocol vulnerabilities, or even the vSwitch software which result in network-based VM attacks.

*Securing VM boundaries* - VMs have virtual boundaries compared with to physical server ones. VMs that co-exist on the same physical server share the same CPU, Memory, I/O, NIC, and others (i.e. there is no physical isolation among VM resources). Securing VM boundaries is the responsibility of the cloud provider.

*Hypervisor security* - a hypervisor is the "virtualizer" that maps from physical resources to virtualized resources and vice versa. It is the main controller of any access to the physical server resources by VMs. Any compromise of the hypervisor violates the security of the VMs because all VMs operations become traced unencrypted. Hypervisor security is the responsibility of cloud providers and the service provider. In this case, the SP is the company that delivers the hypervisor software such as VMware or Xen.

### B. PaaS Security Issues

*SOA related security issues* – the PaaS model is based on the Service-oriented Architecture (SOA) model. This leads to inheriting all security issues that exist in the SOA domain such as DOS attacks, Man-in-the-middle attacks, XML-related attacks, Replay attacks, Dictionary attacks, Injection attacks and input validation related attacks [9, 16]. Mutual authentication, authorization and WS-Security standards are important to secure the cloud provided services. This security issue is a shared responsibility among cloud providers, service providers and consumers.

*API Security* - PaaS may offer APIs that deliver management functions such as business functions, security functions, application management, etc. Such APIs should be provided with security controls and standards implemented, such as OAuth [17], to enforce consistent authentication and authorization on calls to such APIs. Moreover, there is a need for the isolation of APIs in memory. This issue is under the responsibility of the cloud service provider.

### C. SaaS Security Issues

In the SaaS model enforcing and maintaining security is a shared responsibility among the cloud providers and service providers (software vendors). The SaaS model inherits the security issues discussed in the previous two models as it is built on top of both of them including data security management [11] (data locality, integrity, segregation, access, confidentiality, backups) and network security.

*Web application vulnerability scanning* - web applications to be hosted on the cloud infrastructure should be validated and scanned for vulnerabilities using web application scanners [18]. Such scanners should be up to date with the recently discovered vulnerabilities and attack paths maintained in the National Vulnerability Database (NVD) and the Common Weaknesses Enumeration (CWE) [19]. Web application firewalls should be in place to mitigate existing/discovered vulnerabilities (examining HTTP requests and responses for applications specific vulnerabilities). The ten most critical web applications vulnerabilities in 2010 listed by OWASP [20] are injection, cross site scripting (Input validation) weaknesses.

*Web application security miss-configuration and breaking* - web application security miss-configuration or weaknesses in application-specific security controls is an important issue in SaaS. Security miss-configuration is also very critical with multi-tenancy where each tenant has their own security configurations that may conflict with each other leading to security holes. It is mostly recommended to depend on cloud provider security controls to enforce and manage security in a consistent, dynamic and robust way.

### D. Cloud Management Security Issues

The Cloud Management Layer (CML) is the "microkernel" that can be extended to incorporate and coordinate different components. The CML components include SLA management, service monitoring, billing, elasticity, IaaS, PaaS, SaaS services registry, and security management of the cloud. Such a layer is very critical since any vulnerability or any breach of this layer will result in an adversary having control, like an administrator, over the whole cloud platform. This layer offers a set of APIs and services to be used by client applications to integrate with the cloud platform. This means that the same security issues of the PaaS model apply to the CML layer as well.

### E. Cloud Access Methods Security Issues

Cloud computing is based on exposing resources over the internet. These resources can be accessed through (1) web browsers (HTTP/HTTPS), in case of web applications - SaaS; (2) SOAP, REST and RPC Protocols, in case of web services and APIs – PaaS and CML APIs; (3) remote connections, VPN and FTP in case of VMs and storage services – IaaS. Security controls should target vulnerabilities related to these protocols to protect data transferred between the cloud platform and the consumers.

## VIII. CLOUD COMPUTING SECURITY ENABLERS

### A. Identity & Access Management (IAM) and Federation

Identity is a core of any security aware system. It allows the users, services, servers, clouds, and any other entities to be recognized by systems and other parties. Identity consists of a set of information associated with a specific entity. This information is relevant based on context. Identity should not disclose user personal information "privacy". Cloud platforms should deliver or support a robust and consistent Identity management system. This system should cover all cloud objects and cloud users with corresponding identity context information. It should include: Identity Provisioning and de-provisioning, identity information privacy, identity linking, identity mapping, identity federation, identity attributes federation, single sign on, authentication and authorization. Such system should adopt existing standards, *such as SPML, SAML, OAuth, and XACML*, to securely federate identities among interacting entities within different domains and cloud platforms.

### B. Key Management

Confidentiality is one of key objectives of the cloud computing security (CIA triad). Encryption is the main solution to the confidentiality objective, for data, processes and communications. Encryption algorithms either symmetric key-based or asymmetric are key-based. Both encryption approaches have a major problem related to encryption key management i.e. how to securely generate, store, access and exchange secrete keys. Moreover, PaaS requires application keys for all APIs and service calls from other applications. The applications' keys must be maintained securely along with all other credentials required by the application to be able to access such APIs.

### C. Security Management

Based on the huge number of cloud stakeholders, the deep dependency stack, and the large number of security controls to deliver security requirements, the cloud security management becomes a more complicated research problem. Security management needs to include security requirements and policies specifications, security controls configurations according to the policies specified, and feedback from the environment and security controls to the security management and the cloud stakeholders. Security management should function as a plug-in for CML.

### D. Secure Software Development Lifecycle

The secure software development lifecycle (SDLC with security engineering activities) includes elicitation of the security requirements, threat modeling, augmentation of security requirements to the systems models and the generated code consequently. The cloud based applications will involve revolution in the lifecycles and tools used to build secure systems. The PaaS provides a set of reusable security enabling components to help developing secured cloud-based applications. Also security engineering of the cloud-based application should change to meet new security requirements imposed on such systems. Applications should support adaptive security (avoiding hardcoded security) to be able to meet vast range of consumers' security requirements. Adaptive application security is based on externalizing/delegating the security enforcement and applications security management to the cloud security management, cloud security services and security controls.

### E. Security-Performance tradeoff optimization

The cloud computing model is based on delivering services using SLAs. SLAs should cover objectives related to performance, reliability, and security. SLAs also define penalties that will be applied in case of SLA violation. Delivering high security level, *as one of SLA objectives*, means consuming much more resources that impact on the performance objective (the more adopted security tools and mechanism, the worst the impact on the performance of the underlying services). Cloud management should consider the trade-off between security and performance using utility functions for security and performance (least security unless stated otherwise). Moreover, we should focus on delivering adaptive security where security controls configurations are based on the current and expected threat level and considering other tradeoffs.

### F. Federation of security among multi-clouds

When a consumer uses applications that depend on services from different clouds, he will need to maintain his security requirements enforced on both clouds and in between. The same case when multiple clouds integrate together to deliver a bigger pool of resources or integrated services, their security requirements needs to be federated and enforced on different involved cloud platforms.

## IX. CONCLUSION

The cloud computing model is one of the promising computing models for service providers, cloud providers and cloud consumers. But to best utilize the model we need to block the existing security holes. Based on the details explained above, we can summarize the cloud security problem as follows:

- Some of the security problems are inherited from the used technologies such as virtualization and SOA.
- Multi-tenancy and isolation is a major dimension in the cloud security problem that requires a vertical solution from the SaaS layer down to physical infrastructure (to develop physical alike boundaries among tenants instead of virtual boundaries currently applied).
- Security management is very critical to control and manage this number of requirements and controls.
- The cloud model should have a holistic security wrapper, *as shown in figure 3*, such that any access to any object of the cloud platform should pass through security components first.

Based on this discussion we recommend that cloud computing security solutions should:

- Focus on the problem abstraction, using model-based approaches to capture different security views and link such views in a holistic cloud security model.
- Inherent in the cloud architecture. Where delivered mechanisms (such as elasticity engines) and APIs should provide flexible security interfaces.
- Support for: multi-tenancy where each user can see only his security configurations, elasticity, to scale up and down based on the current context.
- Support integration and coordination with other security controls at different layers to deliver integrated security.
- Be adaptive to meet continuous environment changes and stakeholders needs.

## X. FUTURE WORK

We are investigating in the cloud security management problem. Our objective is to block the hole arise in the security management processes of the cloud consumers and the cloud providers from adopting the cloud model. To be able to resolve such problem we need to: (1) Capture different stakeholders security requirements from different perspectives and different levels of details; (2) Map security requirements to the cloud architecture, security patterns and security enforcement mechanisms; and (3) Deliver feedback about the current security status to the cloud providers and consumers. We propose to adopt an adaptive model-based approach in tackling the cloud security management problem. Models will help in the problem abstraction and the capturing of security requirements of different stakeholders at different levels of details. Adaptive-ness will help in delivering an integrated, dynamic and enforceable cloud security model. The feedback loop will measure the security status to help improving the current cloud security model and keeping cloud consumers aware with their assets' security status (applying the trust but verify concept).

## ACKNOWLEDGEMENTS

We thank Swinburne University of Technology for support for parts of this research.